\documentstyle[twoside,fleqn,espcrc2,epsf,amssymb,amsfonts]{article}

\title{Inflation from String/M-Theory Compactification$?$
\thanks{Based on work
done in collaborations with Chiang-Mei Chen, Pei-Ming Ho, John
Wang, and Nobuyoshi Ohta. To appear in Nucl. Phys. B Proceedings
Supplement of Lattice 2003 (Gravity), Tsukuba, Japan, July 2003.}
\author{Ishwaree~P.~Neupane
\address{Department of Physics,
National Taiwan University, 106 Taipei, Taiwan, R.O.C.}}
\\}

\begin{document}

\begin{abstract}

We present some exact scalar potentials for the dimensionally
reduced theory and examine the possibility of obtaining
accelerating $4d$ cosmology from String/M-theory, more generally,
hyperbolic and flux compactification. In the hyperbolic case, even
in the zero-flux limit, the scalar potential is positive for the
$4d$ effective theory as required to get an accelerating universe,
and thereby evading the ``no-go theorem'' given for static
internal space. When we turn on the gauge fields as source terms
at the cosmological background with potential $V\propto {\rm
e}^{-2c\varphi}$, we find eternally accelerating cosmologies when
the $4d$ space-time is flat and $c\leq 1$, or hyperbolic and
$1<c<\sqrt{2}$.

\vspace{0mm}
\end{abstract}

\maketitle

The recent interest in time-dependent String or M-theory
background via space-like branes is two folds. One is the
hypothesized dS/CFT correspondence where one would like to know
whether there can exist a positive extremum of the scalar
potential in a time-dependent String/M-theory compactification.
Another is the consensus that $4d$ cosmology should be derived
from (compactifiaction of) String/M-theory which at low energies
are described by supergravities. This is also motivated from the
recent observation that the present universe is undergoing
accelerated expansion, as indicated by the results from Type Ia
supernovae and CMB measurement by WMAP. Many effective models may
be devised to explain the cosmic acceleration. A more fundamental
task is to derive the $4d$ cosmology from some fundamental theory
of gravity, such as, String or M-theory.

The obstacle for a de Sitter-type compactification in String
theory was ``no-go theorem''~\cite{GWG}. Given that the internal
space is time-independent, and the strong energy condition
$R_{00}\geq 0$ holds for $10$ or $11d$ String or M-thoery, then it
holds also for a compactified theory. The condition
$R_{00}^{(4)}=-\frac{3\ddot{a}}{a} \geq 0$ appears to forbid the
acceleration of the $4d$ spacetime. If so, the low energy SUGRA
limit of superstring theory may not give accelerating FLRW
universe from compactification. However, recently, it has been
shown that it is possible to explain the cosmic acceleration of
$4d$ space-time from supergravity vacuum
solutions~\cite{Townsend03a} or S-brane solutions~\cite{NO2}, if
one gives up the condition of time-independence of internal space,
and, in addition, the internal space is hyperbolic (that is, a
space having the negative scalar curvature). In the subsequent
work~\cite{SRoy,EG,CHNW03a,CHNOW03b,GKL}, the idea of hyperbolic
extra dimensions was found interesting enough to warrant further
investigations of some of its cosmological implications, e.g., the
cosmic acceleration of the present universe.

It is not difficult to understand why hyperbolic extra dimensions
can give rise to a cosmic acceleration of $4d$ space-time. This is
an immediate effect of the positive potentials generated by flux
and hyperbolic compactifications. To be more clear, consider the
bosonic part of D-dimensional supergravity with a $(q+2)$-form
field strength
\begin{equation} {\cal L}_D = \frac{1}{16\pi G_{D}} \sqrt{-g_D}
\left( R - \frac{8\pi G_D}{(q+2)!}\, F_{[q+2]}^2\right)\,,
\end{equation}
where $F_{(q+2)}=dA_{(q+1)}$. We take the metric ansatz in the
Einstein conformal frame~\cite{CHNW03a}
\begin{eqnarray}
& ds_D^2= e^{-\frac{2m}{d-2}\,\phi(t)} \Big[-dt^2+a(t)^2\nonumber \\
&\times\left(\frac{dr^2}{1-\epsilon_0 r^2}+ r^2
d\Omega_{d-2}^2\right)\Big]+ r_1^2\,e^{2\phi(t)} d\Sigma_{m,
\epsilon_1}^2 \,,
\end{eqnarray}
and $d=q+2$. The values of $\epsilon_i=-1,\,0,\,+1$ correspond to
the hyperbolic, flat or spherical space. In $D=11$ (i.e., $d=4$,
$m=7$), one has $4$-form anti-symmetric tensor matter fields as
required by supersymmetry. One uses the $3$-form potential
$A_{abc}$ to preferentially split a space-like three-dimensional
manifold ${\cal M}_3$ from the remaining seven-dimensional space
${\cal M}_7$. $A_{abc}$ is actually required to live (i.e., to be
a maximally form-invariant tensor) on ${\cal M}_3$ or ${\cal
M}_7$. When we take
$$A_{abc} = \sqrt{g_3} \epsilon_{abc}\, {A}(\tau) $$
the ansatz for the field strength is $\ast F _{[q+2]}= 2b \,{\rm
vol}(\Sigma_{\epsilon_1,m})$, $b$ is the field strength parameter.
Upon the dimensional reduction, we find that the effective
Lagrangian is
\begin{equation}\label{action1}
\label{reduction1} {\cal L}_d ={M_{d}^2}\, \sqrt{-g_d}\, \left(
\frac{R_{(d)}}{2}-\Lambda_d+ K-V(\phi)\right)\,,
\end{equation}
where the kinetic and potential terms are
\begin{eqnarray}
&K= -\,\frac{\lambda}{2}\, g^{\mu\nu}\partial_\mu \phi
\partial_\nu\phi\,,\label{kinetic1}\\
&\Lambda_d= -\,\epsilon_0\,\frac{(d-2)(d-3)}{2a^2}\,
\label{curve-poten} \\
&V(\phi)= b^2e^{-\frac{2(d-1)m}{d-2}\phi}-
\frac{m(m-1)\epsilon_1}{2 r_1^2}{\rm
e}^{-\frac{2\lambda}{m}\phi(t)}, \label{M-poten1}
\end{eqnarray}
with $\lambda\equiv \frac{m(m+d-2)}{(d-2)}$. The total potential
$V$ is split into $\Lambda_d$ and $V(\varphi)$; $\Lambda_d$ comes
from the $d$-dimensional space-time curvature. Of course,
$\Lambda_2=\Lambda_3=0$ even if $\epsilon_0\neq 0$. It turns out
that the negatively curved geometry of the internal space (i.e.,
$\epsilon_1=-1$) gives a positive exponential potential $V(\phi)$
in $d$-dimensions~\cite{EG,CHNW03a,CHNOW03b}. Indeed, from an
effective $4d$ theory viewpoint, a scalar potential which is
positive in all (or some) regions of field space is a must to get
an accelerating universe from String/M-theory compactification.

Consider that the $4d$ space-time is flat ($\epsilon_0=0$), and
the internal space is a product of two or more non-trivial curved
spaces of dimensions $m_1,\,m_2,\cdots m_n$, with $m=\sum_i^n
{m_i}$, $\phi_i=\phi_i(t)$. For $b=0$, the kinetic and potential
terms are
\begin{eqnarray}
& K = \frac{1}{2}\sum_i \lambda_i \dot{\phi_i}^2 +
\frac{1}{2}\sum_{i>j=1}^{n} {m_i m_j} {\dot\phi}_i
{\dot\phi}_j\,,  \\
&\Lambda_4=0, \quad V(\phi)= \beta_i\, {\rm
e}^{-2\frac{\lambda_i}{m_i}\phi_i} {\rm e}^{-\sum_{j\neq i}^{1\leq
j\leq n} m_j \phi_j },
\end{eqnarray}
where $\lambda_i=\frac{m_i(m_i+2)}{2}$,
$\beta_i=-\sum_{i}\epsilon_i\,\frac{m_i(m_i-1)}{2\,r_i^2}$, and
$\dot{\phi}=d\phi/dt$. The kinetic term can easily be diagonalized
and normalized by a field re-definition. In the simplest case of
two non-trivial internal spaces $\Sigma_{m_1,\epsilon_1}$,
$\Sigma_{m_2,\epsilon_2}$, we have
\begin{equation}\label{fieldeqs2}
K=\frac{1}{2}\sum_{i=i}^{2}{\dot\varphi}_i^2\,, \quad
V(\varphi)=\beta_i e^{-2\sum_{j=1}^{2}\alpha_{ij}\varphi_j}\,,
\end{equation}
where
\begin{eqnarray}
&\varphi_1=p\phi_1+ \frac{m_1m_2}{2p}\phi_2\,, \quad \varphi_2=
q\phi_2 \,,\\
& \alpha_{ij}= \left(\begin{array}{cc}
p/m_1 & 0 \\
m_1/p & q/m_2
\end{array} \right)\,,
\\
& p\equiv \sqrt{\frac{m_1(m_1+2)}{2}}\,, \quad q\equiv
\sqrt{\frac{m_2(m_1+m_2+2)}{m_1+2}} \,.
\end{eqnarray}
To the $4d$ field equations derived for the kinetic and potential
terms given in~(\ref{fieldeqs2}), we can find exact solutions only
in the two cases: when the internal space is a product of (i) two
similar spaces, e.g., $\epsilon_1=\epsilon_2=-1$, and (ii) flat
and hyperbolic (or spherical) spaces, e.g., $\epsilon_1=0$,
$\epsilon_2=-1$. In fact, if we assume $r_1=r_2$, the solutions
will be further restricted, in which case we require $m_1=m_2$.

For a single internal space, from the action~(\ref{action1}) we
find that the $\varphi$ equation of motion is
\begin{equation}\label{waveeqn}
c\left(\ddot{\varphi}+3H\dot{\varphi}\right)+2\epsilon_1 c^2\,{\rm
e}^{-2c\,\varphi} -3\,\tilde{b}^2 {\rm e}^{-(6/c)\varphi}=0\,,
\end{equation}
while the Friedman equation is
\begin{equation}\label{Friedman}
3H^2=\dot{\varphi}^2-2\epsilon_1 {\rm
e}^{-2c\,\varphi}-\frac{3\epsilon_0}{a^2}+\tilde{b}^2 {\rm
e}^{-(6/c)\varphi}\,,\end{equation} where $c\equiv
\sqrt{\frac{m+2}{m}}$ and $H=\dot{a}(t)/a(t)$. The above $\varphi$
is related to the original $\phi$ by the relation
\begin{equation}
\varphi=\sqrt{\frac{\lambda}{2}}\,\phi-\frac{1}{c} \ln
\frac{\sqrt{m(m-1)}}{2}
\end{equation}
and $\tilde{b}=b\left(4/m(m-1)\right)^{3/2c^2}$.

Let us first consider the case where the external space is flat
(i.e., $\epsilon_0=0$) and the internal space is hyperbolic (i.e.,
$\epsilon_1=-1$), and $b=0$. In this case, one finds convenient to
define
\begin{equation}
d\tau={\rm e}^{-c\,\varphi}\, dt\,, \quad
\alpha(\tau)=\ln(a(t))\,.
\end{equation}
A class of accelerating solutions as implied by the equations
(\ref{waveeqn}), (\ref{Friedman}), (see also
Ref.~\cite{Townsend03a}), is
\begin{eqnarray}
&{\alpha^\prime}=\frac{d\alpha(\tau)}{d\tau}
=\frac{2\lambda_-\left(6\lambda_+\cosh^2(\gamma
\tau)-1\right)\gamma}{\sinh(2\gamma \tau)}\,,\\
 &{\rm
e}^{-\varphi}=\left(\cosh\gamma\tau\right)^{-\sqrt{3}\lambda_-}
\left(\sinh\gamma\tau\right)^{\sqrt{3}\lambda_+}\,,
\end{eqnarray}
\begin{equation}
\lambda_{\pm} = \frac{1}{\sqrt{3}(\sqrt{3}\pm c)}\,, \quad
\gamma=\sqrt{\frac{1}{6\lambda_+\lambda_-}}\,.
\end{equation}
These yield, for $\tau>0$, in the units $r_1=1$, 
\begin{eqnarray}
&H=\frac{da/dt}{a}= {\rm e}^{-c\varphi}\,\alpha^\prime(\tau)>0
\,,\\
&\frac{e^{2c\varphi}}{2\gamma^2}
\frac{\ddot{a}}{a}=\frac{2(c^2-1)}{c^2-3}
+ \frac{2\sqrt{3}c \left(2\cosh^2\gamma
\tau-1\right)-c^2-3}{3(3-c^2) \left(\cosh^2\gamma
\tau-1\right)\cosh^2\gamma \tau}\,.
\end{eqnarray}
Although only $c>1$ arises in hyperbolic compactification, one may
consider the critical value $c=1$, which separates qualitatively
the different cosmologies. Likewise, $c=\sqrt{3}$, so-called
`hyper-critical' (see~\cite{Townsend03a} and references therein)
separates hyperbolic from flux compactification, which is a
special case, and practically we take here $1< c<\sqrt{3}$.
Interestingly, when $c=1$ and $\tau>0.2746$, the solution is
always accelerating (there also exists an eternally decelerating
universe for $\tau<-0.2746$). While, the solution with $c>1$ is
only transiently accelerating.

For the compactification of $11$-dimensional supergravity on a
$7$-dimensional compact hyperbolic space, one has $c=3/\sqrt{7}$,
and hence the condition for acceleration ($\ddot{a}/a>0$) is
satisfied in a (symmetric) interval
$$
0.4296\lesssim |\tau| \lesssim 1.4025 $$ However, only for an
interval of positive $\tau$, the $4d$ space-time is (transiently)
accelerating. Note that in the interval $-1.4025<\tau <-0.4296$,
the universe is contracting since $\dot{a}/a<0$.

Turn to the case of $\epsilon_0=\epsilon_1=-1$. In the zero-flux
limit, it is easy to find the exact solution~\cite{CHNOW03b}
\begin{equation}
a=\frac{c}{\sqrt{c^2-1}}\,t\equiv a_0\,,\quad \varphi=
\frac{1}{c}\ln(ct)\equiv \varphi_0\,.
\end{equation}
This solution itself is not accelerating since $\ddot{a}(t)=0$,
but, to the lowest order, where a non-zero field strength
parameter $b>0$ serves as a source term, the solution is
accelerating when $c<\sqrt{2}$. For example, when $c=\sqrt{4/3}$,
we find
\begin{equation}
a(t)\sim a_0+\frac{0.78\,\tilde{b}^2}{t^{3/2}}\,,\quad
\varphi\sim\varphi_0+\frac{0.74\,\tilde{b}^2}{t^{5/2}} \,.
\end{equation}
One may verify numerically, without turning to perturbation, that
the system of equations (\ref{waveeqn}), (\ref{Friedman}) gives an
eternally accelerating expansion. The existence of a period of
acceleration in a $\epsilon_0=-1$, $\epsilon_1=0$ cosmology was
noted before~\cite{Costa02a}.

We have derived some exact scalar potentials for the dimensionally
reduced theory, analyze some simple exponential potentials from
the viewpoint of $4d$ cosmology, and examine the possibility of
generating inflation/acceleration from flux and hyperbolic
compactifications, in general. To most ansatzs for metric
decomposition, the acceleration of $4d$ space-time is transient,
and leads to only a few e-foldings. However, in the presence of
matter fields, this may be implemented to explain the late-time
inflation of the universe, which may require some fine-tuning
among dark energy density, mass of the massive Kaluza-Klein modes
and effective dimensionality of the universe.

A more promising aspect of the hyperbolic compactification is
that, in a non-trivial form-field background, an exponential
potential $V\propto e^{-2c\varphi}$, with $1 \leq c<\sqrt{2}$, can
give eternally accelerating cosmologies.

\section*{Acknowledgment}
\noindent I wish to thank Chiang-Mei Chen, Pei-Ming Ho, Nobuyoshi
Ohta and John Wang for fruitful collaborations and Paul Townsend
for helpful correspondences. I am grateful to the organizers of
the Lattice 2003 and NCTS, Taiwan for their generous support.

\end{document}